\documentstyle{article}
\begin{document}
\title{\Large  Relationship Between Structural Fractal and Possible
                  Dynamic Scaling Properties in Protein Folding  \\}
\author{\large Liang-Jian Zou and X. G. Gong  \\
  Institute of Solid State Physics, Academia Sinica,
  P.O.Box 1129, Hefei 230031, China        \\
  CCAST (World Laboratory), P.O.Box 8730, Beijing, 100080,China\\
  International Center for Theoretical Physics, ICTP, P.O.Box 586,
  34100, Trieste, Italy \\
  Zhu Zheng-Gang \\
  Laboratory of Internal Friction and Defects in Solids,
  Institute of Solid State Physics,\\ Academia Sinica,
  P.O.Box 1129, Hefei 230031, China        \\}
\date{  }
\maketitle
\pagenumbering{arabic}
\large
\begin{center}
{\bf  Abstract} \\
\end{center}
   In this letter, the possible dynamic scaling properties of protein molecules
in folding are investigated theoretically by assuming that the protein
molecules are percolated networks. It is shown that the fractal character and
the fractal dimensionality may exist only for short sequences in large protein
molecules and small protein molecules
with homogeneous structure, the fractal dimensionality are obtained for
different structures. We then show that there might exist the dynamic scaling
properties in protein folding,
the critical exponents in the folding for some small
global proteins with homogeneous structure are obtained. The dynamic critical 
exponents of the global proteins in folding are relevant to the fractal
dimensionality of its structure, which implies the close relationship
between the dynamic process in protein folding and its structure kinematics.

\vspace{1cm}


\vspace{1cm}
\newpage
\large
   The prediction for the compact spatial structure of folded protein and the
folding process of extended protein molecule has attracted much attention
in this decade, since it is realized that the unique, native conformation
of protein molecule has close relationship with its biophysical functions.
The aim of the study on the protein folding is trying to
understand why and how an extended protein folds into its unique and 
native state quickly through 
the astronomic number possible intermediate states. Though many efforts
have afforded for it and the progress has been made step by step, it is clear
that we are still far 
away from our goal to elucidate the folding process in detail [ 1, 2 ].

    The difficulty of the study on the structure prediction of folded protein 
and the folding process comes from several aspects: (1) the complexity of
the constitutions, a protein molecule may contain 20 kinds amino-acids, (2)
the randomness of the sequences and
the complexity of the structure (including $\alpha$-helix, $\beta$-sheet,
$\tau$-turn, etc.) and (3) the giant atom-molecule assemble of the
protein-solvent system. In the past years, the study on the protein folding
and the structure prediction is
mainly in the atom-molecule level, it is based on the interaction between
atoms or amino-acid resides, such as by the molecular dynamic simulation method
[ 3 - 4 ] relying on the empirical potential between the
atoms or the amino-acids and the (lattice) Monte Carlo simulation 
method [ 5 - 8 ]. These
methods could provide the detail process of folding and the final
stable state of protein, however, in dealing with large protein molecules,
these methods may meet difficulty arising from both the limit of computer
and the accuracy of the empirical potential. On the other hand, some authors
tried to understand the protein structure in whole scale [ 9 - 10 ].
The plot of the spatial structure for the backbone of protein molecule
shows that the backbone could be
achieved by Brownian motion, or self-avoiding random walk. The fractal
character of protein molecule is thus one of the most attracting aspect.
Stapleton [ 9 ] studied the spectral dimensionality of myoglobin and some 
other proteins by the electron spin-relaxation measurements, and after then
some authors tried to classify the proteins in terms of the fractal
dimensionality (FD). However, a series subsequent
investigations [ 10 ] later show that it is difficult to define
a general FD for a large protein molecule because of its nonhomogeneous
structure and the absence of the complete self-similarity. It is clear 
that the FD for helix and for sheet structures
are different, so characterizing a general protein molecule
containing both helix and sheet structures by its fractal and classifying
proteins in terms of its FD are not easy.

   In the past few years, some authors [ 11 - 17 ] experimentally found that
the protein solution may exhibit the critical phenomena, and several critical
exponents for the protein and water solution are obtained. However,
little is known theoretically for such phenomena and the global dynamics of
the whole protein in folding. In this letter,
we will illustrate that the FD of certain short sequences of large
and small protein moleculse with the homogeneous structure
can be defined approximately, though it is difficult to define an unique and
unified FD for a large and complicated molecule. In the following,
by assuming that the folding 
polypeptide is a percolating network and through the scaling law,
 we can obtain the critical exponents 
in the folding process, and try to reveal some common characters of
different kinds of proteins in folding processes.

   The number of the hydrogen bonds and disulphate bonds is small in an
unfolded or denatured protein, and these bonds distribute
randomly. In the folded state, the number of the hydrogen bonds is large,
and the whole polypeptide chain are connected by those bonds.
One result of the connection is that the extended polypeptide chain becomes
gradually unsmooth, some sequences in the chain may have self-similarity and
exhibit the local fractal characters.
Since FD is the measure of the torsion and the curve for unsmooth lines,
the stronger the torsion of the systems is, the larger the FD is, and vice
visa. Therefore to some extent, the FD does reflect the structure information
of proteins including the second and the tertiary structure, or even more the
quaternary structure.
So the FD may be a natural measure of the folding degree and the local
structure of protein molecule.

 Though it is difficult to define a general FD for a large protein molecule,
detail analysis shows that it is fractal for certain short sequences with
homogeneous structure in large protein molecule. This can be supp orted
by such a fact that starting from one startpoint of the sequence, the
Ln(L)-Ln(r) plot (here r is end-end distance and L the backbone length of
the protein sequence) is approximately a straight line. When
the Ln(L)-Ln(r) plot 
deviates from linear, it suggests that the polypeptide chain or the
sequence begin to twist with itself, the protein molecule chain tends to form
the ternary structure or the quaternary structure. As we will see below, the
FD of protein sequences with different structures are different

  Flavodoxin protein contains 148 amino-acid resides, it is a typical
example in which the concept of "fractal for certain sequence" is adequate.
Fig.1 shows the Ln(r)-N plot and the Ln(L)-Ln(r) plot for flavodoxin
protein, here N denotes the atom number of the backbone from one endpoint.
It can be seen from Fig.1a that for a few sequences of the flavodoxin
molecule, the Ln(r)-N curve is approximately a straight line. In Fig.1b,
for every specific sequence, the slope of the ln(L)-Ln(r)
curve is almost a constant, therefore for every sequence, an approximate
FD can be defined. Also small protein molecules with homogeneous
structure have similar properties.

  For a series of small proteins and short sequences in large proteins with 
the structures well-defined,
their FD are shown in Fig.2, here the well-defined structure means
the structure is homogeneous, the FD is obtained by linear best fit.
Accordingly, the average FD are 1.378$+/-$0.200 for helix sequence and
1.088$+/-$0.020 for sheet sequence, respectively. It is found that for
helix sequence, the longer the sequence is, the smaller the FD is
(See Fig.2a). However the
FD are almost same for different lengths of sheet sequences (See Fig2b). From
the above discussion,
we show that the fractal can be well-defined for small proteins or short
sequences in large proteins with homogeneous structure. Here and below the FD
is referred to that of the small proteins or short
sequences in the large proteins with homogeneous structure.

  As we all know, a protein chain in native state is neither a completely
disordered state nor a completely ordered state since it contains both some
regular structures ($\alpha$-helix, $\beta$-sheet, etc.) and some irregular
ones. By mapping a hydrogen bonds to a connected state, it is more
suitable to consider a folded protein molecule as a percolated network.
For a polypeptide chain in folding, when and where a hydrogen bond forms
are stochastic. We can consider the polypeptide chain as a
network connected by the hydrogen bonds, the number of the hydrogen bonds
can be described as the percolated degree. Thus
one can define the relative number of the
hydrogen bonds, {\it p}, as an order parameter. At the critical value of the
order parameter, where {\ p}={\it p$_{c}$},
most of the hydrogen bonds form and the protein enters its native state.
When a hydrogen bond or disulphate bond forms, which connects two sites 
nearby or far away in the chain, it is considered that the percolation occurs.

  As pointed out above, the fractal character can be defined for a small 
protein molecule or a
short sequence with homogeneous structure, and the protein chain
can be considered as a percolated network. In this case, the folding
protein may exhibit percolating behaviors, such as the critical
characters.
In fact, during the folding process, the protein molecule behaves highly
cooperatively and like a phase transition, the critical exponents and the
scaling power
of the folding then can be obtained in the percolation theory.
By the scaling argument and the scaling relationship, one can easily obtain
one of the scaling powers and two critical exponents through the FD of the
three-dimension spatial structures of the small protein molecules or short
sequences. One of the scaling powers relating to the hydrogen bonds is:
\begin{equation}
     a_{H} =\frac{d_{f}}{d}
\end{equation}
where d$_{f}$ is the FD of a  protein network, and d the Euclidean
dimensionality in space. This is an interesting
result. The scaling power a$_{H}$, hence the critical exponents, of a
folding protein depend only on its Euclidean dimension and its FD, which
suggests that the dynamics of the protein is determined only by its
global structure. One of the important properties of the protein network
is the correlation between amino-acids or atoms in different
sites. In the folding, the correlation function, $f(|{\bf r}-{\bf r'}|)$,
may exhibit critical behavior, $f(r) \approx r^{-(d-2+\eta)}$. By the scaling
relation, one can obtain the critical exponent of the correlation
function between sites in the same chain, $\eta$:
\begin{equation}
    \eta =2+d-d_{f}
\end{equation}
Another one of the important properties of the protein network
is its behavior of the "free energy" of the whole molecule, or the G function,
also it may exhibit critical character, $G \approx (p-p_{c})^{\delta}$.
Through the scaling relation,
the critical exponent for the G function is:
\begin{equation}
    \delta =\frac{d_{f}}{d-d_{f}}
\end{equation}
One notices that these two critical exponents depend only on the FD and the
Euclidian dimensionality of the molecule.
It is well-known that the six critical exponents of a percolated network can
be derived from two independent scaling powers.
In the present letter, only one of the two independent scaling powers is
determined, the another one needs further study.

 From the preceding discussion, one can relate the structure kinematics of a
protein to the dynamic scaling behavior through the FD. Accordingly,
the scaling powers a$_{H}$ are about 0.460 for helix and 0.363 for sheet
structures, respectively. The critical
exponents for the correlation function of different sites in the same chain 
are 3.62 for helix and 3.91 for sheet structures, respectively. The
critical exponents for the G function of the hydrogen bonds are about
0.85 for helix sequences and 0.569 for sheet sequences, respectively.
Obviously, for a protein chain containing both the helix and the sheet
structures, the scaling power and the critical exponents should lie between
these values.

   The relationship between the FD and dynamic scaling properties has its
physical origin. As we all know, the physical force field determines the
configuration and the conformation of the protein chain, and the dynamic
scaling thus depends on the interaction of the atoms in the protein chain.

It should stressed here that the present results is only adequate
for the dynamics of single protein
molecule. One way of the measurement for
the dynamic scaling behavior in the folding is to measure the correlation of
different sites by the neutron scattering experiments, it may give the data
of the critical exponents of the correlation function in some protein 
molecules.
Also we notice that the present theory doesn't consider the influence of the
water environment, so it is difficult to compare the present theoretical
results with the available experimental data of protein-water solution
[ 11 - 17 ]. However, 
many of the protein and other biological molecules in aqueous solution
may interconnect through the hydrogen bonds, so the protein-water systems may
behave like a huge percolated network. Some results developed here might
be suitable for such systems.

 In summary, the relationship between the structural fractal and possible
dynamic scaling properties in protein folding is explored. It is found that 
the
FD may be well-defined for homogeneous protein structures, and the folded
protein can be regarded as a percolated network. One of the scaling powers is
found to depend only on the FD and the Euclidean dimensionality. The critical
exponents for the correlation function and the G-function are obtained. 
Although
these theoretical results are obtained for complicated protein molecules,
clearly it could be applied for the homologous
polymers.            \\

 ACKNOWLEDGEMENT: One of the authors (L.-J. Zou) thanks Prof. Yu Lu and the
invitation of the International Center for Theoretical Physics (ICTP) in 
Trieste. This work is financially supported by the Grant of
National Natural Science Foundation of China No.19477104 and in part by the
Fund of National Laboratory of Internal Friction and Defects in Solids.

\newpage
\large
\vspace{1.5cm}

\begin{center}
{\bf REFERENCES }
\end{center}

\begin{enumerate}

\item F. M. Richards,{\it Protein Folding}, Ed. by T. E. Creighton, Chapt.1,
      (W.H.Freeman and Company, New York 1992)

\item F. M. Richards and W. A. Lin, Quartly Review of Biophysics, {\bf 26} 
      423, (1994);

\item M. Karplus,  {\it Structural Molecular Biology}, Ed. by D. B.
      Davies,  (Plenum, New York 1981)

\item W. F. Von Gunsteren, et al., Proc. Natl. Acad. Sci. USA,
      biophys. {\bf 80}, 4315 (1983)

\item P. E. Leopold, M. Montal, and J. N. Onuchic, Proc. Natl. Acad. Sci. 
      USA, {\it biophys}. {\bf 89}, 8721 (1992)

\item E. I. Shakhnovich,  Phys. Rev. Lett., {\bf 72}, 3907 (1994)

\item A. Wallqvist and M. Ullner, Protein: {\it Structure, Function and
      Genetics}, {\bf 18}, 267 (1994)

\item A. Kolinski and J. Skolnick, Protein: {\it Structure, Function and
      Genetics}, {\bf 18}, 338, 353 (1994)

\item H. J. Stapleton, J. P. Allen., C. P. Flynn, D. G. Stinson and S. R.
      Kurtz,  Phys. Rev. Lett., {\bf 45}, 1456 (1980)

\item Li Hou-Qiang and Wang Fu-Quan, {\it Fractal Theory and its Application}
      {\it in Molecular Science}, Chapt. 7-8, ( Scientific Press, Beijing, 
      1993) and some reference therein.

\item B. M. Fine, J. Pande, A. Lowakin, O. O. Ogun and G. B.
      Benedek,  Phys. Rev. Lett., {\bf 74}, 198 (1995)

\item E. I. Shakhnovich and A. M. Gutin,   Proc. Natl. Acad. Sci. USA,
      {\bf 90}, 7195 (1993)

\item J. J. Ramsden  Phys. Rev. Lett., {\bf 71}, 295 (1993)

\item C. Ishimoto and T. Tanaka,  Phys. Rev. Lett., {\bf 79}, 474 (1977)

\item P. Schurtenberger, R. A. Chanberlin, G. M. Thurston, J. A. Thomson and
      G. B. Benedek,  Phys. Rev. Lett., {\bf 63}, 2064 (1989)

\item J. A. Thomson, P. Schurtenberger, G. M. Thurston, and G. B. Benedek,
      Proc. Natl. Acad. Sci. USA, {\it biophysics} {\bf 84}, 7079 (1987)

\item M. L. Broide, C. R. Berland, J. Pande, O. O. Ogun and B. Benedek,
      Proc. Natl. Acad. Sci. USA, {\it biophysics} {\bf 88}, 5660 (1991)

\end{enumerate}

\newpage
\begin{center}
{\bf  Figures Captions }
\end{center}
\vspace{1cm}
\noindent Fig. 1. The dependence of the end-end distance (r) of flavodoxin
protein on the number of the sequences (N) and the length of chain (L).
(a). Ln(r) vs. N plot, and (b). Ln(r) vs. ln(L) plot.  \\

\vspace{1cm}

\noindent Fig. 2. The fractal dimensionality of sheet structure for 84
           protein sequences (a) and helix structure for 182
           sequences (b) with homogeneous structure.
           The average FD is 1.081 for sheet structure (a) and.
           1.378 (b)for helix structure, respectively. \\

\vspace{1cm}
\end{document}